\documentclass[12pt]{article}
\usepackage{epsfig}
\usepackage{latexsym}
\begin{document}
\title{Lorentz invariant quantization of the Yang-Mills theory free of Gribov ambiguity.}
\author{ A.A.Slavnov \thanks{E-mail:$~~$ slavnov@mi.ras.ru}
\\Steklov Mathematical Institute, Russian Academy
of Sciences\\ Gubkina st.8, GSP-1,119991, Moscow} \maketitle

\begin{abstract}
A new formulation of the Yang-Mills theory which allows to avoid the problem of Gribov ambiguity of the gauge fixing is proposed.
\end{abstract}

\section {Introduction}

The standard formulation of the Yang-Mills theory does not allow a unique gauge fixing. It was shown by V.N.Gribov \cite{Gr} that the Coulomb gauge condition $\partial_iA_i=0$ does not choose a unique representative in the class of gauge equivalent configurations, as the condition 
\begin{equation}
\partial_iA^{\Omega}_i=0
\label{1}
\end{equation}
 considered as the equation for the elements of the gauge group $\Omega$ at the 
 surface $\partial_iA^i=0$ for sufficiently large $A$ has nontrivial solutions fastly decreasing at the spatial infinity. This result was generalised by I.Singer \cite{Si} to arbitrary covariant gauge conditions.

In the framework of perturbation theory, that is for sufficiently small $A$ the equaton (\ref{1}) has only trivial solutions. Hence the Gribov umbiguity in this case is absent. However beyond the perturbation theory this ambiguity exists, which makes problematic the standard way of canonical quantization of nonabelian gauge theories. This problem was studied by many authors (see e.g. \cite{Zw}), but in my opinion the problem is still far from being clear.

Recently I proposed an explicitely Lorentz invariant formulation of the quantum Yang-Mills theory in which the effective Lagrangian of ghost fields is gauge invariant \cite{Sl1}. In the present paper I will show that in this approach the Yang-Mills theory allows a quantization procedure which is free of the Gribov ambiguity and hence may serve as a starting point for nonperturbative constructions.

\section{Unambiguos quantization of the Yang-Mills field.}

We consider the model described by the classical Lagrangian
\begin{equation}
L=- \frac{1}{4}F_{\mu\nu}^aF_{\mu\nu}^a+(D_{\mu} \varphi)^*(D_{\mu} \varphi)-(D_{\mu}\chi)^*(D_{\mu}\chi)+ i[(D_{\mu}b)^*(D_{\mu}e)-(D_{\mu}e)^*(D_{\mu}b)]
\label{2}
\end{equation}
To save the place we shall consider the model with the $SU(2)$ gauge group. Generalization to other groups makes no problem. Here $F_{\mu\nu}^a$ is the standard curvature tensor for the Yang-Mills field. The scalar fields $\varphi, \chi, b, e$ form the complex $SU(2)$ doublets parametrized by the hermitean components as follows:
\begin{equation} 
 c=\left( \frac{ic_1+c_2}{\sqrt{2}}, \frac{c_0-ic_3}{\sqrt{2}}\right)
\label{3}
\end{equation}
where $c$ denotes any of doublets. The fields $\varphi$ and $\chi$ are commuting, and the fields $e$ and $b$ are anticommuting. In the eq.(\ref{2}) $D_{\mu}$ denotes the usual covariant derivative, hence the Lagrangian (\ref{2}) is gauge invariant. Note that due to the negative sign of the $\chi$ field Lagrangian, this field posesses negative energy. 

Let us make the following shifts in the Lagrangian (\ref{2}):
\begin{equation}
\varphi \rightarrow \varphi +g^{-1}\hat{m}; \quad \chi \rightarrow \chi -g^{-1}\hat{m}; \quad \hat{m}=(0,m)
\label{4}
\end{equation}
where $m$ is a constant parameter. Due to the negative sign of the Lagrangian of the field $\chi$ the terms quadratic in $m$ arising due to the shifts of the fields $\varphi$ and $\chi$ mutualy compensate and the Lagrangian acquires a form
\begin{eqnarray}
L=- \frac{1}{4}F_{\mu\nu}^aF_{\mu\nu}^a+(D_{\mu}\varphi)^*(D_{\mu}\varphi)-(D_{\mu}\chi)^*
(D_{\mu}\chi)\nonumber\\ +g^{-1}[(D_{\mu}\varphi)^*+(D_{\mu}\chi)^*](D_{\mu}\hat{m})+g^{-1}(D_{\mu}\hat{m})^*[D_{\mu}\varphi+D_{\mu}\chi]\nonumber\\ +i[(D_{\mu}b)^*(D_{\mu}e)-(D_{\mu}e)^*(D_{\mu}b)]
\label{5}
\end{eqnarray}
As before this Lagrangian describes massless vector particles.

The Lagrangian (\ref{5}) is obviously invariant with respect to the "`shifted"' gauge transformations, which in terms of hermitean components look as follows
\begin {eqnarray}
\delta A^a_{\mu}= \partial_{\mu}\eta^a-g\epsilon^{abc}A^b_{\mu}\eta^c\nonumber\\
\delta \varphi_0^+= \frac{g}{2} \varphi^+_a \eta^a\nonumber\\
\delta \varphi_0^-=\frac{g}{2} \varphi^-_a \eta^a\nonumber\\
\delta \varphi^+_a=-\frac{g}{2} \epsilon^{abc}\varphi^+_b\eta^c- \frac{g}{2} \varphi^+_0 \eta^a\nonumber\\	
\delta \varphi^-_a=-m\eta^a-\frac{g}{2} \epsilon^{abc}\varphi^-_b \eta^c-\frac{g}{2} \varphi^-_0\eta^a\nonumber\\
\delta b^a=-\frac{g}{2}  \epsilon^{adc}b^d \eta^c-\frac{g}{2}b^0\eta^a\nonumber\\
\delta e^a=-\frac{g}{2} \epsilon^{adc}e^d\eta^c-\frac{g}{2}e^0 \eta^a\nonumber\\
\delta b^0= \frac{g}{2}b^a \eta^a\nonumber\\  
\delta e^0= \frac{g}{2}e^a \eta^a
\label{6}
\end{eqnarray}

Here the obvious notations 
$$ 
\varphi_{\alpha}^{ \pm}= \frac{\varphi_{\alpha} \pm \chi_{\alpha}}{\sqrt{2}}
$$
are introduced.

The Lagrangian (\ref{5}) except for gauge invariance posesses also the invariance with respect to the supersymmetry transformations
\begin{eqnarray}
\delta \varphi(x)=i \epsilon b(x)\nonumber\\
\delta \chi(x)=-i \epsilon b(x)\nonumber\\
\delta e(x)=\epsilon [\varphi(x)+ \chi(x)]\nonumber\\
\delta b(x)=0
\label{7}
\end{eqnarray}
where $\epsilon$ is a constant anticommuting parameter.

In the future we shall see that invariance with respect to the supersymmetry transformations provides unitarity of the theory in the space which includes only physical exitations of the fields. An explicit form of interaction is not essential. Only the symmetry properties are important. In principle any counterterms which preserve gauge invariance and supersymmetry are allowed.

To quantize the model one has to impose a gauge condition. One could choose the Coulomb gauge, as it was done in our paper \cite{Sl1}. The supersymmetry transformations do not involve the gauge fields, hence the Coulomb gauge Lagrangian preserves the supersymmetry. However in this case we would confront again the problem of Gribov ambiguity. Although in the framework of perturbation theory such a choice does not lead to any problems and allows to prove easily equivalence of the model (\ref{5}) to the usual Yang-Mills theory, beyond the peturbation theory the ambiguity problem may be essential.

The explicit form of the gauge transformations (\ref{6}) shows that under these transformations not only the Yang-Mills fields are shifted by the gradients of arbitrary functions, but also the fields $\varphi^-_a$ are shifted by arbitrary functions. It allows to add to the Lagrangian (\ref{5}) the gauge fixing term of the form
\begin{equation}
\frac{1}{2 \alpha}(\varphi^-_a)^2
\label{8}
\end{equation}
and in particular to consider the gauge
\begin{equation}
\varphi^-_a=0
\label{9}
\end{equation}
Note that the relation
\begin{equation}
(\varphi^\Omega)^-_a=0
\label{9a}
\end{equation}
considered as the equation for the group elements $\Omega$ at the surface (\ref{9}) has no notrivial solutions fastly decreasing at spatial infinity. So imposing the gauge (\ref{9}) is unambigous.

In the gauge $\varphi^-_a=0$ the action acquires a form
\begin{eqnarray}
\tilde{A}= \int d^4x \{- \frac{1}{4}F_{\mu\nu}^aF_{\mu\nu}^a+ \partial_{\mu}\varphi^+_0 \partial_{\mu}\varphi^-_0 +m \varphi^+_a \partial_{\mu}A_{\mu}^a\nonumber\\  +i[(D_{\mu}b)^*(D_{\mu}e)-(D_{\mu}e)^*(D_{\mu}b)]\nonumber\\+ \frac{mg}{4}A_{\mu}^2\varphi_0^++ \frac{g^2}{8}A_{\mu}^2\varphi_0^+\varphi_0^-+g \partial_{\mu}\varphi_0^-A_{\mu}^a\varphi_a^++ \frac{g}{2}\varphi_0^-\varphi_a^+ \partial_{\mu}A_{\mu}^a \}
\label{10}
\end{eqnarray}

The canonical momentum for the field $A_0^a$ is
\begin{equation}
p_0^a=m \varphi_a^+(1+ \frac{g}{2m}\varphi_0^-)
\label{11}
\end{equation}
The Hamiltonian action looks as follows
\begin{eqnarray}
\tilde{A}_H= \int d^4x \{p_i^a \dot{A_i^a}+p_0^a \dot{A_0^a}+p_{\varphi} \dot{\varphi_0}+p_{\chi} \dot{\chi_0}\nonumber\\ - \frac{(p_i^a)^2}{2}+ \frac{(p_0^a)^2}{2(1+g/(2m)\varphi_0^-)^2}- \frac{p_{\varphi}^2}{2}+ \frac{p_{\chi}^2}{2}+ \nonumber\\
+A_0^a \partial_ip_i^a- \frac{p_0^a \partial_iA_i^a}{1+g/(2m)\varphi^-_0}- \frac{1}{4}F^a_{ik}F^a_{ik}+ \ldots \}
\label{12}
\end{eqnarray}
Here $\ldots$ denote the terms corresponding to the fields $b$ and $e$ and all the interaction terms.
 We wrote explicitely the term quadratic in momenta $p_0^a$ because it generates a nontrivial Jacobian when one passes from the path integral over the phase space to the integral in the coordinate space. Due to the presence of this factor integration over canonical momenta generates the additional term in the measure
\begin{equation}
\prod_x(1+ \frac{g}{2m}\varphi_0^-)^3
\label{13}
\end{equation}
and the scattering matrix in the gauge $\varphi_a^-=0$ may be written as a path integral
\begin{equation}
S= \int \exp\left\{i\tilde{A}\right\}d \mu
\label{14}
\end{equation}
where
\begin{equation}
d \mu=\prod_x(m+ \frac{g}{2}\varphi_0^-)^3 dA_{\mu}d \varphi_{\alpha}^+
d \varphi_0^-d b^{\alpha}d e^{\alpha}
\label{15}
\end{equation}

The scattering matrix (\ref{14}) acts in the space which contains unobservable exitations: unphysical components of $A_{\mu}$, ghosts corresponding to the fields $\varphi_0^+, \varphi_0^-,b_{\alpha}, e_{\alpha}$. Below we shall show that the Lagrangian (\ref{5}) in the gauge $\varphi_a^-=0$ similar to the Lagrangian in the Coulomb gauge is invariant under some supersymmetry transformations which provide the unitarity of the scattering matrix in the space including only physical exitations of the Yang-Mills field.

\section {Physical unitarity of the model.}

In classical theory the transition from the Coulomb gauge to the gauge $\varphi^-_a=0$ in the Lagrangian (\ref{5}) may be done with the help of the gauge transformation (\ref{6}). A gauge transformation may be considered as a change of variables which does not alter the action. In the Coulomb gauge the action was invariant with respect to the supersymmetry transformation (\ref{7}). Hence the action in the gauge $\varphi^-_a=0$ also will be ivariant under some supersymmetry transformation. However this transformation differs of (\ref{7}). The explicit form of this transformation may be found as follows. Under a gauge transformation the action does not change, but the gauge fixing term $ \int d^4x \lambda^a(x) \partial_iA^a_i(x)$ in new variables will be replaced by $ \int d^4x \lambda^a(x) \tilde{\varphi}^-_a(x)$. In this way we get the equation determining the gauge function
\begin{equation}
\int d^4x\lambda^a(x)\partial_i(A_\Omega)_i^a(x)= \int d^4x \lambda^a(x)\varphi^-_a(x)
\label{16}
\end{equation}
Both gauges under consideration are admissible, therefore the equation (\ref{16}) has a solution. In practice one can find a solution of the eq.(\ref{16}) in perturbation theory. At the lowest order we have
\begin{equation}
\eta^a(x)= \frac{-\varphi^-_a(x)+ \partial_iA_i^a(x)}{m}
\label{17}
\end{equation}
where the functions $\eta^a(x)$ parametrize the gauge group elements $\Omega(x)$.
Under the supersymmetry transfomations the function $\eta$ changes
\begin{equation}
\eta^a(x) \rightarrow \eta^a(x)-i \frac{\sqrt{2} \epsilon b^a(x)}{m}
\label{18}
\end{equation}
Therefore
\begin{eqnarray}
\tilde{A}_{\mu}^a(x)=A_{\mu}^a(x)- \partial_{\mu}\eta^a(x) \rightarrow  \tilde{A}_{\mu}^a(x)+i \frac{\sqrt{2} \epsilon \partial_{\mu}\tilde{b}^a(x)}{m}\nonumber\\
\tilde{e}_{\alpha}(x) \rightarrow \tilde{e}_{\alpha}(x)+ \epsilon \sqrt{2}\tilde{\varphi}_{\alpha}^+(x)\nonumber\\
\tilde{\varphi}^-_0(x) \rightarrow \tilde{\varphi}^-_0(x)+i \sqrt{2} \epsilon \tilde{b}_0(x)
\label{19}
\end{eqnarray}
The remaining fields do not change at zero order.

Invariance with respect to the supersymmetry transformations which in the free theory are described by the eqs.(\ref{19}), according to the Noether theorem generates a conserved charge $Q$. For asymptotic fields this charge reduces to the free one $Q_0$, which is determined by the transformations (\ref{19}). Note that here we rely on the hypothesis about adiabatic switching of interaction for asymptotic states. Validity of this hypothesis for nonabelian gauge fields is not obvious. For this reason a formal proof given below should be applied either to infrared regularized theory \cite{Sl2} or to the functional generating Green functions of gauge invariant composite operators.

Physical states are separated by the condition
\begin{equation}
Q_0|\psi>_{as}=0
\label{20}
\end{equation}
The explicit form of the charge $Q_0$ may be obtained by considering variation of the free Lagrangian under the transformations (\ref{19}). It may be presented in the form
\begin{equation}
Q_0= \tilde{Q}_0^0+Q_0^0
\label{20a}
\end{equation}
where
\begin{equation}
\tilde{Q}^0_0=\sqrt{2} \int d^3x \{m^{-1}( \partial_iA_0- \partial_0A_i)^a(\partial_ib^a)-\varphi^+_a \partial_0b^a \}
\label{20b}
\end{equation}
and
\begin{equation}
Q^0_0= \sqrt{2} \int d^3x \{ \partial_0 \varphi^+_0 b_0+ \partial_0b_0 \varphi^+_0 \} 
\label{21}
\end{equation}
Here the operator $\tilde{Q}_0^0$ coincides with the free BRST-operator for the Yang-Mills theory in the Lorentz gauge if one identifies $b_a$ with the Faddeev-Popov ghost $c_a$ and $e_a$ with $\bar{c}_a$.
The operator $Q_0^0$ in terms of creation and annihilation operators looks as follows
\begin{equation}
Q_0^0 \sim \int d^3k  \{ b_0^*(\textbf{k})\varphi_+^0(\textbf{k})+(\varphi_+^{0})^*(\textbf{k})b_0(\textbf{k}) \}
\label{22}
\end{equation}
where
\begin{equation}
\left[\varphi^-_0(\textbf{k}), \varphi^{+*}_0(\textbf{p})\right]=\delta(\textbf{k}-\textbf{p})
\label{24}
\end{equation}
\begin{equation}
[e_0(\textbf{k}),b^*_0(\textbf{p})]_+=\delta(\textbf{k}-\textbf{p})
\label{25}
\end{equation}
The operators $\tilde{Q}_0^0$ and $Q_0^0$ are independent and anticommuting. So the equation (\ref{21}) may be fulfilled only if the admissible vectors $|\psi>_{as}$ are annihilated by both operators $\tilde{Q}_0^0, \quad Q_0^0$.

As it is well known from the theory of BRST-quantization(see \cite{KO}, any vector annihilated by $\tilde{Q}_0^0$ may be presented in the form
\begin{equation}
|\psi>= | \tilde{\psi}>+ |\tilde{N}>
\label{25a}
\end{equation}
where the vector $| \tilde{\psi}>$ contains only the exitations corresponding to three dimensionally transversal components of the Yang-Mills field and the particles corresponding to $ \varphi_0^+, \varphi_0^-,b_0, e_o$, and $|\tilde{N}>$ is a zero norm vector.

Further proof of nonnegativity of the norms of the vectors satisfying the condition (\ref{20}) goes in a standard way \cite{He}. We introduce the number operator for unphysical particles corresponding to the operators $e_0, b_0, \varphi^+_0, \varphi^-_0$
\begin{equation}\hat{N}= \int d^3k \{b^*_0e_0+e^*_0b_0+ \varphi^{+*}_0 \varphi^-_0+\varphi^{-*}_0 \varphi_0^+ \} 
\label{26}
\end{equation}
and notice that this operator may be presented as the anticommutator 
\begin{equation}
\hat{N}=[Q_0^0,K_0]_+
\label{27}
\end{equation}
where
\begin{equation}
K_0=\int d^3k \{e_0^*(\textbf{k})\varphi_0^-(\textbf{k})+\varphi_0^{-*}(\textbf{k})e_0(\textbf{k}) \}
\label{28}
\end{equation}
Any vector with $N$ unphysical particles $\varphi^{\pm}_0, b_0, e_0; \quad (N \neq 0)$ looks as follows
\begin{equation}
|\psi>= \frac{1}{N} \{Q_0^0K_0|\psi>+K_0Q_0^0|\psi>\}
\label{29}
\end{equation}
If this vector is annihilated by the operator $Q_0^0$ 
\begin{equation}
|\psi>= \frac{1}{N}Q_0^0|\chi>
\label{30}
\end{equation}
Combining the equations (\ref{25a}, \ref{30}) one sees that any vector annihilated by the operator $Q_0$ may be presented as 
\begin{equation}
|\psi>_{as}=|\psi>_{tr}+|N>
\label{31}
\end{equation}
where the vector $|\psi>_{tr}$ contains only exitations corresponding to the three dimensionally transversal components of the field $A_{\mu}$, and the vector $|N>$ has zero norm. Factorizing this space with respect to zero norm vectors we see that it coincides with the "physical" space of the Yang-Mills theory. By construction the scattering matrix is unitary in this space, and expectation value of any gauge invariant operator over states (\ref{20}) coincides with the expectation value over transversal state vectors.

The expression (\ref{14}) for the scattering matrix in principle makes sense beyound perturbation theory. Similar expression may be written for the functional generating Green functions of gauge invariant composite operators. 

For practical calculations of such Green functions in the framework of perturbation theory it is convenient to pass in the path integral to the Lorentz gauge. It can be done if one notes that at the surface $\varphi_a^-=0$ the factor $\prod_x(a+ \frac{g}{2} \varphi_0^-)^3$ may be written in the gauge invariant form
\begin{equation}
\prod_x(a+ \frac{g}{2} \varphi_0^-)^{-3}= \int d\Omega \delta(\varphi^{-\Omega}_a)|_{\varphi_a^-=0}
\label{32}
\end{equation}
where the integration at the r.h.s. goes over invariant measure at the gauge group.

Using this observation one can write the expression for the gauge invariant generating functional in the form
\begin{equation}
Z(J_{\mu})= \int \exp \{i(A+ \int d^4x J_{\mu}F_{\mu})\}\delta(\varphi_a^-)\Delta_-dA_{\mu}d \varphi_{\alpha}^-\varphi_{\alpha}^+db_{\alpha}de_{\alpha}
\label{33}
\end{equation}
where $F_{\mu}$ denotes a gauge invariant functional, $A$ is the gauge invariant action, corresponding to the Lagrangian (\ref{5}) and 
\begin{equation}
\Delta_- \int d\Omega \delta(\partial_{\mu}A_{\mu}^\Omega)=1
\label{34}
\end{equation}
Multiplying the integral (\ref{33}) by "`one"
\begin{equation}
\Delta_L \int \delta(\partial_{\mu}A_{\mu}^\Omega)d\Omega=1
\label{35}
\end{equation}
and making the change of variables which is the gauge transformation (\ref{6}) we obtain the expression for the generating functional in the Lorentz gauge
\begin{equation}
Z(J_{\mu}= \int \exp\left\{i(A+ \int d^4xJ_{\mu}F_{\mu})\right\}\Delta_L\delta(\partial_{\mu}A_{\mu})dA_{\mu}d \varphi_{\alpha}db_{\alpha}de_{\alpha}
\label{36}
\end{equation}
In the framework of perturbation theory the transformation to the Lorentz gauge is well defined. The perturbation series in the Lorentz gauge is explicitely renormalizable.

One can also develop the perturbation theory directly in the gauge $\varphi_a^-=0$. However the propagator of the fields $\varphi_a^+(x),A_{\mu}^b(y)$ at large momenta behaves as $k^{-1}$ which generates in the perturbation theory an infinite number of types of divergent diagrams. Nevertheless the degree of divergency of all these diagrams is limited, which allows to hope on the renormalizability of the model in this gauge as well. (An example of such situation is given by the supersymmetric gauge models in manifestly supersymmetric gauges). The explicit calculations of the $\beta$-function at the lowest order (\cite{Ba}) confirms this hypothesis. The model under consideration is asymptotically free and the running charge is expressed by the usual formula.

\section{Discussion}

In the present paper we constructed the quantization procedure for nonabelian gauge theories which allows to get a unique expression for the Green functions of gauge invariant composite operators and infrared regularized scattering matrix. Imposing the gauge condition does not introduce the ambiguity indicated by Gribov. Lorentz invariance of the theory is also manifest. In the framework of perturbation theory this model leads to the expression for the gauge invariant Green functions, which coincides with the usual one.

 {\bf Acknowledgements.} \\ This work was partially done while the author was
 visiting University of Milan. I wish to thank R.Ferrari
  for hospitality and Cariplo Foundation for a
 generous support. My thanks to the members of Theoretical group for
 helpful discussion. This researsh was supported in part by Russian
 Basic Research Fund under grant 08-01-00281a and by the RAS program "`Nonlinear dynamics".$$ ~ $$
  \end{document}